\documentclass[aps,amsmath,amssymb,nofootinbib,twocolumn]{revtex4}
\usepackage{graphicx}
\usepackage{pgfplots}
\pgfplotsset{compat=1.17}
\usepackage{dcolumn}
\usepackage{color}
\usepackage{scrextend}
\usepackage{subfig}
\usepackage{bm}
\usepackage{amsmath,amssymb,graphicx}
\usepackage{epsfig}
\usepackage{enumerate}
\usepackage{array}
\usepackage{multirow}
\usepackage{tikz}
\usepackage{ragged2e}
\usepackage{textgreek}
\usepackage[normalem]{ulem}
\usepackage{mathtools}

\begin{document}
\title{Modeling the non-Markovian Brownian\\ motion of an optomechanical resonator}
\author{Aritra Ghosh$^1$\footnote{aritraghosh500@gmail.com}, Malay Bandyopadhyay$^2$, and M. Bhattacharya$^1$}
\affiliation{$^1$School of Physics and Astronomy, Rochester Institute of Technology, 84 Lomb Memorial Drive, Rochester, New York 14623, USA\\
$^2$School of Basic Sciences, Indian Institute of Technology Bhubaneswar, Argul, Jatni, Khurda, Odisha 752050, India}
\vskip-2.8cm
\date{\today}
\vskip-0.9cm

\vspace{5mm}
\begin{abstract}
We propose a representative, globally-admissible phenomenological spectral density of the bath for the non-Markovian Brownian motion of an optomechanical resonator, motivated by the near-resonance experimental observation of a non-Ohmic spectrum in [Nat. Commun. \textbf{6}, 7606 (2015)]. To avoid divergences arising from a naive global extrapolation, we propose a globally-admissible, phenomenological bath spectrum that extends the experimentally-observed, near-resonance non-Ohmic behavior beyond the measurement window while ensuring finite bath-induced renormalizations and quadrature fluctuations of the resonator. The corresponding model of the structured environment produces a nonlocal mechanical response whose analytic pole structure encodes the observed linewidth. The resulting dissipation kernel exhibits a power-law-modulated exponential decay with transient negativity, signaling memory effects. In the weak-coupling regime, the optical readout based on homodyne detection enables near-resonance spectroscopy and, with a calibrated drive on the resonator, permits, in principle, the reconstruction of the full mechanical susceptibility, thereby providing access to both the dissipative and dispersive bath contributions. Our results provide a consistent route from locally-inferred spectral properties to globally-admissible open-system descriptions and establish a framework for probing structured environments in cavity optomechanics.
\end{abstract}

\maketitle

\section{Introduction}
Interactions between a system and its environment are ubiquitous in nature and play a central role in determining the dynamics of quantum systems \cite{Weiss_2021,Breuer_Petruccione_2002}. In the weak-damping paradigm, this environmental interaction is routinely simplified using the Markov approximation, assuming a memoryless environment. However, for micro- and nanomechanical systems in solid-state architectures, the environment is often fundamentally structured. Mechanisms such as coupling between flexural modes and two-level systems \cite{Seoanez_2007}, or clamping losses and phonon tunneling \cite{Wilson-Rae_2008}, can give rise to a non-Ohmic environmental spectral density. As a result, the system undergoes non-Markovian quantum Brownian motion \cite{Grabert_1988,Hanggi_2005,Ghosh_2024}, resulting in time-delayed damping and colored noisy fluctuations. Probing these effects requires an experimental platform of exquisite sensitivity. Over the past few years, cavity optomechanics \cite{Kippenberg_2007,Aspelmeyer_2014}, exploiting the radiation-pressure coupling between the optical cavity field and mechanical vibrations, has emerged as an important platform for this purpose \cite{Groeblacher_2015,Jiang_2020,Zhang_2021}. By allowing monitoring of the transmitted light, optomechanical systems can act as sensitive probes of non-Markovianity. 

\vspace{2mm}

A central pertinent question is how the information encoded in such optical signatures can be promoted to a physically-consistent characterization of the mechanical environment. While Gr\"oblacher \textit{et al.} \cite{Groeblacher_2015} have experimentally inferred the local frequency dependence of the environmental spectral density in a narrow window around the mechanical resonance, finding a sub-Ohmic slope $k = -2.30 \pm 1.05$ that departs significantly from the Ohmic profile, a general framework that connects this local spectroscopic observation to a globally-defined bath model remains absent. This issue is important for both fundamental and practical reasons. The environmental spectral density governs not only dissipation but also memory effects, colored fluctuations, and the bath-induced renormalization of the mechanical response. So without a globally-admissible description of the bath, one remains largely confined to a narrow frequency window around the resonance. This motivates the present work where our aim is to bridge local evidence of non-Markovianity and global bath characterization by constructing a nonunique phenomenological spectral density that is compatible with the experimentally-inferred non-Ohmic behavior. In particular, we shall relate the locally-observed, non-Ohmic behavior directly to a globally-consistent description of the mechanical environment.

\vspace{2mm}

Beyond establishing a consistent bath model, we shall further develop a spectroscopic framework formulated directly at the level of the mechanical susceptibility and bath self-energy. Unlike \cite{Zhang_2021}, where the bath spectral function was inferred from the optical response, we distinguish between passive measurements and coherent-force spectroscopy, the latter enabling, in principle, a route to reconstructing the full complex mechanical response under appropriate calibration and weak-probe conditions. This formulation provides simultaneous access to both the dissipative and dispersive components of the bath self-energy. Consequently, it provides direct access to the bath self-energy, whose imaginary part determines dissipation while its real part determines frequency renormalization. The proposed framework therefore not only captures the near-resonance behavior observed experimentally but also offers a route to a characterization of structured environments in micromechanical systems.

\vspace{2mm}

The paper is organized as follows. The theoretical model based on the linear-coupling Hamiltonian and its preliminary consequences are presented in Sec. \ref{model_sec}. This is then followed by our extrapolation of the near-resonance information of \cite{Groeblacher_2015} to the entire bath spectrum based on certain consistency requirements in Sec. \ref{model_J_sec}. The position and momentum correlation functions are then analyzed in Sec. \ref{corr_sec}, and the nonuniqueness of the global completion of the spectral density from the local slope is discussed in Sec. \ref{nonunique_sec}. Sec. \ref{spec_sec} includes our protocol for probing non-Markovianity in the mechanical motion. The paper is then concluded in Sec. \ref{conc_sec}. Several technical details of the calculations are supplied in the Appendices \ref{appA}-\ref{appF}. 

\section{Theoretical model}\label{model_sec}
\begin{figure}
\centering
\includegraphics[width=1\linewidth]{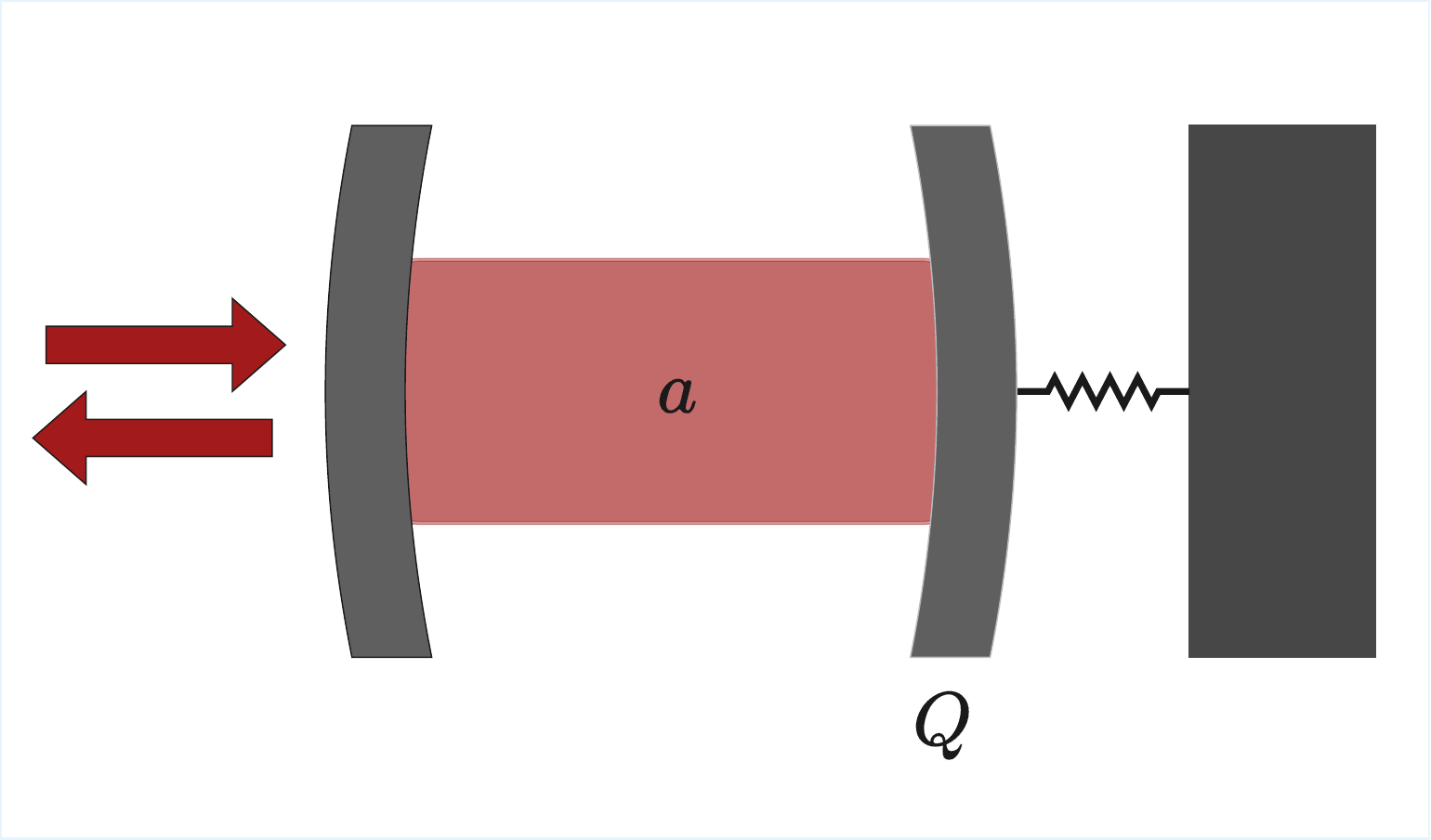}
\caption{\justifying{Schematic of an optomechanical setup where a control laser enters the cavity from the left mirror and the intracavity mode $a$ couples with the micromechanical motion of the right mirror, represented by the operator $Q$ denoting the displacement of the resonator's center-of-mass position from its mean value. Assuming negligible intrinsic losses, the input and output fields are indicated with the arrows.}}
\label{fig1}
\end{figure}

The system of interest is an optomechanical resonator, shown schematically in Fig. \ref{fig1}. Considering only the mechanical resonator along with the environment, it can be described by the linear-coupling model \cite{Ullersma_1966,Ford_1988}
\begin{equation}
H = \frac{P^2}{2M} +\frac{1}{2}M\Omega_0^2 Q^2 +
\sum_{j=1}^N \left[\frac{p_j^2}{2m_j} +\frac{1}{2}m_j\omega_j^2 x_j^2\right] - Q\sum_{j=1}^N c_j x_j,
\label{H}
\end{equation} where $M$ and $\Omega_0$ are the bare mass and frequency of the resonator. The heat bath is modeled as a collection of independent quantum oscillators \cite{Feynman_1963,Ford_1965,Caldeira_1983} of masses $\{m_j\}$ and frequencies $\{\omega_j\}$, with $j=1,2,\cdots,N$; $N$ here being a large number. The system-bath couplings $\{c_j\}$ are taken to be real, and the operators satisfy the standard commutation relations (we shall work with $\hbar=1$)
\begin{equation}
    [Q,P] = i, \quad \quad [x_j,p_k] = i\delta_{j,k},
\end{equation}
with all others vanishing. With these notations the bath spectral function is defined as \cite{Weiss_2021}
\begin{eqnarray}
J(\omega)&=&\frac{\pi}{2}\sum_j \frac{c_j^2}{m_j\omega_j}\delta(\omega-\omega_j) \nonumber \\
&\simeq&\frac{\pi}{2} \rho(\omega) \frac{c(\omega)^2}{m(\omega) \omega},
\label{J_discrete}
\end{eqnarray} 
where the last asymptotic equality holds in the continuum limit of the heat bath, with bath density of states $\rho(\omega)$, coupling function $c(\omega)$, and mass profile $m(\omega)$. The system's position operator satisfies a quantum Langevin equation \cite{Ford_1988,Ghosh_2024} (see Appendix \ref{appA})
\begin{equation}\label{QLE}
M \ddot{Q} + \int_{-\infty}^t \mu(t-t') \dot{Q}(t') dt' + \left(M\Omega_0^2-\delta K\right)Q(t) = F(t),
\end{equation}
where $F(t)$ is the thermal noise and one encounters a nonlocal damping, characterized by the kernel $\mu(t)$ and related to the bath spectral function as \cite{Weiss_2021}
\begin{equation}
\mu(t) = \frac{2}{\pi} \int_0^\infty \frac{J(\omega)}{\omega} \cos (\omega t) d\omega,
\label{mut_J}
\end{equation} with $t \geq 0$ for causality. The noise correlations $C_{FF}(t-t') = \frac{1}{2}\left\langle
\left\{F(t),F(t')\right\}\right\rangle$ are 
\begin{equation}
C_{FF}(t-t') = \frac{1}{\pi}\int_0^\infty d\omega J(\omega) 
\coth \left(\frac{\omega}{2 k_B T}\right)
\cos\bigl[\omega(t-t')\bigr].
\label{noise_correlations}
\end{equation}
Within the present Gaussian microscopic model, the specification of the bath spectral function $J(\omega)$ supplies all the relevant information concerning the heat bath at a given temperature. Notably, we have omitted the usual Caldeira-Leggett counter-term \cite{Caldeira_1983} (see also, \cite{Ford_1988}) in writing the Hamiltonian operator in Eq. (\ref{H}) in order to be fully consistent with the treatment of \cite{Groeblacher_2015}. This naturally leads to a bath-induced shift in the resonator's stiffness, as given by
\begin{equation}
\delta K = \mu(0) = \frac{2}{\pi}\int_0^\infty d\omega\frac{J(\omega)}{\omega}. 
\label{deltaK_def}
\end{equation}
Our goal is not to keep the bare frequency $\Omega_0$ fixed but to treat the experimentally-observed resonance $\Omega_R$ (to be defined later) as the physical frequency after dressing by the bath \cite{Weiss_2021,Grabert_1988}. Now the quantum Langevin equation [Eq. (\ref{QLE})] can be solved in the Fourier domain as $\tilde{Q}(\omega) = \chi(\omega) \tilde{F}(\omega)$, where `tilde' denotes Fourier transform in the convention $\tilde{o}(\omega) = \int_{-\infty}^{\infty} o(t) e^{i\omega t} dt$. The susceptibility is given by
\begin{equation}
\chi^{-1}(\omega) = M \Omega_0^2 - M \omega^2 - \delta K - \Sigma(\omega),
\label{chi_general}
\end{equation}
where $\Sigma(\omega)$ denotes the regularized self-energy due to the bath, being defined in our convention as
\begin{equation}\label{principal_value_exp}
\Sigma(\omega) = \frac{2}{\pi}\mathcal{P}\int_0^\infty d\omega' J(\omega') \left[\frac{\omega'}{\omega'^2-\omega^2} - \frac{1}{\omega'}\right] + iJ(\omega).
\end{equation}
In Eq. (\ref{principal_value_exp}) $J(\omega)$ is defined for $\omega>0$. When required, one can employ the retarded continuation to the full real axis as ${\rm Im}[\Sigma^R(\omega)]={\rm sgn}(\omega)J(|\omega|)$, so that $\Sigma^R(-\omega)=[\Sigma^R(\omega)]^*$ and $\chi(-\omega)=\chi^*(\omega)$. It is useful to isolate the quadratic low-frequency contribution when it exists. Eq. (\ref{principal_value_exp}) can equivalently be written as
\begin{equation}
{\rm Re}[\Sigma(\omega)]
=
\frac{2\omega^2}{\pi}\mathcal{P}\int_0^\infty
d\omega' \frac{J(\omega')}{\omega'(\omega'^2-\omega^2)}.
\end{equation}
For the smooth spectra considered here, provided
$J(\omega')/\omega'^3$ is integrable in the infrared and ultraviolet, one has
\begin{equation}
\lim_{\omega\to0} \frac{{\rm Re}[\Sigma(\omega)]}{\omega^2} = \delta M, \quad \quad
\delta M= \frac{2}{\pi}\int_0^\infty d\omega'\frac{J(\omega')}{\omega'^3}.
\label{deltaM_def}
\end{equation}
Thus ${\rm Re}[\Sigma(\omega)]=\delta M\omega^2+o(\omega^2)$. Let us write
$\Sigma(\omega)=\delta M\omega^2+\Sigma_{\rm res}(\omega)$, where $\Sigma_{\rm res}(\omega)$ contains the remaining dispersive contribution together with the dissipative term $iJ(\omega)$. Substituting this into Eq. (\ref{chi_general}) yields
\begin{equation}
\chi^{-1}(\omega) = \bigl(M\Omega_0^2-\delta K\bigr) - \bigl(M+\delta M\bigr)\omega^2 - \Sigma_{\rm res}(\omega).
\label{chi_ren}
\end{equation}
One can view $\delta K$ and $\delta M$ as bath-induced renormalizations of the resonator's stiffness and mass \cite{Grabert_1988}, and the present description is physically meaningful if $M\Omega_0^2 - \delta K > 0$, in which case the bath-induced stiffness renormalization does not make the resonator unstable.

\section{Model bath spectrum from near-resonance estimate}\label{model_J_sec}
In this section we shall construct a bath spectral function that is globally defined and consistent with the experimental observation of its near-resonance behavior $J_k(\omega)|_{\omega \approx \Omega_R} \propto \omega^k$, where $k = -2.30 \pm 1.05$ \cite{Groeblacher_2015}. It is noteworthy that the experimental exponent $k$ should be regarded only as a local characterization of the environment in the vicinity of the resonance frequency $\Omega_R$. That is, it does not by itself determine the full bath spectral function over all frequencies. Our subsequent construction is therefore intentionally extrapolative: it embeds the measured local slope into a spectral density that is globally defined, which can then be used for analytical calculations.

\vspace{2mm}

Let us begin by first characterizing the near-resonance behavior of the effective description. In the weak-damping regime \cite{Grabert_1984,Karrlein_1997}, the observed mechanical resonance is well approximated by the frequency $\Omega_R$ satisfying ${\rm Re}[\chi^{-1}(\Omega_R)]=0$, giving the relation
\begin{equation}
M\Omega_0^2-\delta K = (M + \delta M)\Omega_R^2+{\rm Re}[\Sigma_{\rm res}(\Omega_R)],
\label{pole_condition}
\end{equation}
where $\Omega_R$ is introduced as an experimentally-specified frequency for the local response theory. More strictly, the observed spectral maximum is determined by the minimum of $|\chi(\omega)|^{-2}$ that also includes the imaginary part ${\rm Im}[\chi^{-1}(\Omega_R)]$. Focusing on a weak-damping regime, the linewidth is narrow and the dissipative part varies only weakly across the resonance window. The observed resonance frequency is therefore well approximated by $\Omega_R$ defined through ${\rm Re}[\chi^{-1}(\Omega_R)]=0$. For this near-resonance description, it is natural to expand the full inverse susceptibility locally about $\omega=\Omega_R$. Introducing
\begin{equation}
M_R = M+ \delta M + \left.\frac{\partial {\rm Re}[\Sigma_{\rm res}(\omega)]}{\partial(\omega^2)}\right|_{\omega=\Omega_R},
\end{equation}
the inverse susceptibility takes the local form
\begin{equation}
\chi_{\rm eff}^{-1}(\omega) \approx M_R(\Omega_R^2-\omega^2)-iJ_k(\Omega_R),
\label{chi_eff}
\end{equation}
to the leading order around the resonance, taking $J_k(\omega) \approx J_k(\Omega_R)$, justified in the weak-damping regime. Assuming $M_R > 0$, this form provides the appropriate starting point for the near-resonance description in which the observed resonance frequency $\Omega_R$ is taken as the physical mechanical frequency.

\subsection{Physically-admissible spectral densities}
Although one does not exactly know the bath spectral function except for its near-resonance scaling, one can impose consistency conditions to find a globally-defined extension that conforms to physical constraints. The first requirement is, of course, the non-negativity of $J_k(\omega)$. Second, for the globally-regular subclass considered here, we shall impose the additional sufficient requirements that both $\delta K$ and the low-frequency quadratic coefficient $\delta M$ be finite. These conditions ensure a finite static stiffness shift and a well-defined quadratic low-frequency inertial renormalization. A further physical requirement is that the stationary fluctuations or variances of the resonator's position and momentum should remain finite. These thermal variances can be calculated by invoking the Callen-Welton form of the fluctuation-dissipation theorem at thermal equilibrium \cite{Weiss_2021}, and as shown in Appendix \ref{appB}, their finiteness is automatic if $\delta M$ and $\delta K$ are well defined. Eqs. (\ref{deltaK_def}) and (\ref{deltaM_def}) show immediately that a globally-imposed power law $J_k(\omega)|_{\omega \approx \Omega_R}\propto \omega^k$ with negative $k$ is unacceptable, as it leads to infrared divergences in the renormalization integrals. Therefore, if a measured local behavior $J_k(\omega)|_{\omega \approx \Omega_R}\propto \omega^k$ is observed only near the resonance, one must embed it into a spectral density that is globally well behaved. In addition, stability requires the bath-renormalized static stiffness to remain positive, i.e., $M\Omega_0^2-\delta K>0$ from Eq. (\ref{pole_condition}). To summarize, we intend to satisfy the following sufficient requirements:

\begin{enumerate}
\item $J_k(\omega)$ must be non-negative. 
\item At the observed mechanical resonance $\Omega_R$, $J_k(\omega)|_{\omega \approx \Omega_R} \propto \omega^k$, with $k = -2.30 \pm 1.05$ \cite{Groeblacher_2015}. 
\item Both $\delta K$ [Eq. (\ref{deltaK_def})] and $\delta M$ [Eq. (\ref{deltaM_def})] should be finite, i.e., nondivergent, with the stability condition $M\Omega_0^2-\delta K>0$. A convenient common and sufficient condition ensuring the ultraviolet and infrared convergence of both the renormalization integrals is that the spectral density should satisfy $J_k(\omega)\sim \mathcal{O}(\omega^s)$ with $s>2$ as $\omega\to0$, and $J_k(\omega)\sim \mathcal{O}(\omega^r)$ with $r<0$ as $\omega\to\infty$.
\end{enumerate}
These sufficient conditions motivate the following class of globally-consistent spectral densities:
\begin{equation}
J_k(\omega)=A_k \omega^{s}f_k\left(\frac{\omega}{\Omega_R}\right),
\end{equation} where $A_k$ is a positive constant depending on $k$, and $f_k(\xi)$ is a dimensionless positive function satisfying $\lim_{\xi\to0}f_k(\xi)=1$ and $f_k(\xi)=\mathcal{O}(\xi^{-(s+\epsilon)})$ as $\xi\to\infty$ for some $\epsilon>0$. Now suppose that we wish to reproduce the prescribed local logarithmic slope
\begin{equation}
\left.\frac{d\ln J}{d\ln\omega}\right|_{\omega=\Omega_R}=k
\end{equation}
at the observed resonance. One then immediately finds
\begin{equation}
s + \left.\frac{d\ln f_k(\xi)}{d\ln \xi }\right|_{\xi=1} = k.
\label{general_slope_condition}
\end{equation}
The general modeling problem is therefore reduced to the choice of an infrared-safe exponent $s$, a positive crossover function $f_k$, and the imposition of the local-slope condition of Eq. (\ref{general_slope_condition}). Throughout, we shall restrict our attention to parameter choices for which the stability condition $M\Omega_0^2-\delta K>0$ holds.

\subsection{Model spectral density}
\begin{figure}
\centering
\includegraphics[width=\linewidth]{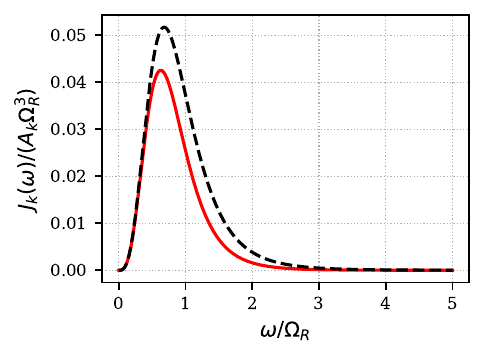}
\caption{\justifying{Normalized bath spectral function $J_k(\omega)/(A_k \Omega_R^3)$ as a function of the dimensionless frequency $\omega/\Omega_R$. The solid-red curve corresponds to the experimental central estimate $k = -2.30$, while the dashed-black curve denotes $k = -1.75$. Both the profiles exhibit a super-Ohmic infrared scaling ($\propto \omega^3$) that guarantees infrared stability of the mass and stiffness renormalizations ($\delta M$ and $\delta K$). The nonmonotonic profile reflects the presence of a structured environment, characterized by a redistribution of the spectral weight around a characteristic frequency scale below the mechanical resonance.}}
\label{fig2}
\end{figure}

To this end, let us propose the following simple form of the bath spectral function:
\begin{equation}
J_k(\omega)=A_k \omega^3\left[1+\left(\frac{\omega}{\Omega_R}\right)^2\right]^{k-3},
\label{J_fit}
\end{equation} 
where we have chosen $s=3$ as the lowest integer satisfying our imposed infrared regularity condition $s>2$. For the experimentally-relevant $k<0$, this spectral density is non-negative, infrared-safe, and ultraviolet-convergent, and it reproduces the target local logarithmic slope exactly at $\omega=\Omega_R$. One can check that in the experimental window $\omega_{\rm min}/2\pi=0.885~{\rm MHz}$ to $\omega_{\rm max}/2\pi=0.945~{\rm MHz}$ around $\Omega_R/2\pi = 0.914~{\rm MHz}$ \cite{Groeblacher_2015}, the logarithmic slope $(d\ln J_k)/(d\ln\omega)$ changes only from $-2.13$ to $-2.48$ for $k=-2.30$, remaining close. Here it is important to re-emphasize that the scale $\Omega_R$ appearing in Eq. (\ref{J_fit}) is treated as the experimentally-observed resonance frequency, not as a quantity to be predicted self-consistently from a microscopic bath model. Thus Eq. (\ref{J_fit}) should be regarded as a phenomenological effective spectral density anchored to the measured resonance scale. Its purpose is to provide a globally-well-defined completion of the locally-observed, near-resonance behavior and not a first-principles bath spectrum from which $\Omega_R$ is independently derived. The prefactor $A_k$ can be inferred from a possible on-resonance determination of the bath spectral function as
\begin{equation}
A_k = 2^{3-k} \frac{J_k(\Omega_R)}{\Omega_R^3}. 
\label{A_k}
\end{equation}
The quantities $\delta K$ and $\delta M$ can now be exactly calculated using this model (see Appendix \ref{appC}). It must be stressed here that the spectral function in Eq. (\ref{J_fit}) is not a unique inference from the experiment but one consistent global extrapolation of the limited spectral information \cite{Groeblacher_2015}. Our choice of $J_k(\omega)$ therefore represents one analytically-tractable completion that preserves the measured local slope while enforcing global consistency.  Moreover, the asymptotic constraints and the condition
$\left.d\ln J/d\ln\omega\right|_{\Omega_R}=k$ leave the intermediate-frequency structure underdetermined. That is, the low- and high-frequency convergence conditions, together with the local slope at $\Omega_R$, do not exclude more complicated intermediate-frequency structures such as additional peaks, dips, or gaps. The present form is therefore applicable when the relevant bath response is smooth over the intermediate-frequency range and is dominated by a single resonance-controlled crossover scale. Other structured cases can be accommodated by choosing a different positive crossover function $f_k(\omega/\Omega_R)$ while retaining the same infrared, ultraviolet, and local-slope constraints. The asymptotics of the present model are
\begin{equation}
J_k(\omega) \sim
\begin{cases}
    A_k \omega^3 & (\omega\ll \Omega_R), \\
    A_k \Omega_R^{2(3-k)}\omega^{2k-3} & (\omega\gg \Omega_R),
\end{cases}
\end{equation}
explicitly confirming the super-Ohmic infrared behavior leading to finite $\delta K$ and $\delta M$. Fig. \ref{fig2} shows the bath spectral function for representative values of $k$, exhibiting nonmonotonic behavior with a maximum at
\begin{equation}
\frac{\omega_{J,{\rm max}}}{\Omega_R} = \sqrt{\frac{3}{3 - 2k}},
\end{equation}
thereby demonstrating that $\omega_{J,{\rm max}} < \Omega_R$ for the experimentally-relevant negative values of $k$. The nonmonotonic behavior of $J_k(\omega)$ is consistent with a structured environment in the sense that dissipation and fluctuations are concentrated around a characteristic frequency scale. 

\vspace{2mm}

An alternate parametrization of the bath characteristics can be obtained from the coupling function $c_k(\omega)$ appearing in Eq. (\ref{J_discrete}), giving the effective strength with which the system coordinate $Q$ couples to the bath modes at frequency $\omega$ in the continuum description. While its particular form is not unique due to explicit dependence on the normalization convention for the bath modes, a representative form in a particular choice of the microscopic modeling can be found by taking $\rho_k(\omega) \simeq \tilde{\rho}$ and $m_k(\omega) \simeq \tilde{m}$ as constants, leading to
\begin{equation}
c_k(\omega) =  \omega^2 \sqrt{\frac{2^{4-k}\tilde{m}J_k(\Omega_R)}{\pi \tilde{\rho}\Omega_R^3}}\left[1+\left(\frac{\omega}{\Omega_R}\right)^2\right]^{\frac{k-3}{2}},
\end{equation}
where we have used Eq. (\ref{A_k}). The coupling function inherits the nonmonotonicity of the bath spectral function though the maximum of $c_k(\omega)$ does not coincide with that of $J_k(\omega)$ and is instead given by $\omega_{c,{\rm max}}/\Omega_R=\sqrt{2/(1-k)}$. This difference can be understood from the physical distinction between the bath spectral function and the coupling function. Since $c_k(\omega)$ describes how strongly the system couples to a single bath mode at frequency $\omega$, while $J_k(\omega)$ measures the effective influence of all modes at that frequency on the system dynamics, $J_k(\omega)$ is shaped not only by the coupling strength but also by how many modes are available and how they are normalized, so its peak does not necessarily occur where the bare coupling $c_k(\omega)$ is the largest. Assuming approximately-constant mode density and modal mass, the observed super-Ohmic infrared behavior can be understood as arising from a low-frequency suppression of the coupling $c_k(\omega)\sim \omega^2$. Our model of the structured mechanical reservoir is thus strongly frequency-selective.

\subsection{Linewidth estimation}
Let us now derive the weak-damping linewidth that governs the experimentally-measured susceptibility. In the near-resonance regime, let us use the local form of the inverse susceptibility given in Eq. (\ref{chi_eff}). Assuming that the response is dominated by an isolated weakly-damped pole, we can write
\begin{equation}
\omega_* \approx \Omega_R-i\frac{\gamma}{2}, \quad \quad
\gamma \ll \Omega_R.
\end{equation}
Expanding to the leading order in $\gamma$ gives
\begin{eqnarray}
\Omega_R^2-\omega_*^2 &=& \Omega_R^2-\left(\Omega_R-i\frac{\gamma}{2}\right)^2 \nonumber \\
&=& i\Omega_R\gamma+\mathcal{O}(\gamma^2).
\end{eqnarray}
Substituting this into the pole condition $\chi_{\rm eff}^{-1}(\omega_*)=0$ yields $iM_R\Omega_R\gamma-iJ_k(\Omega_R)\approx 0$, so that
\begin{equation}
\gamma\approx \frac{J_k(\Omega_R)}{M_R\Omega_R}.
\label{gamma_pole}
\end{equation}
The stability of the pole $(\gamma > 0)$ together with $J_k(\Omega_R) > 0$ conforms to the positivity of $M_R$. This linewidth estimate is valid provided the pole is both weakly damped ($\gamma \ll \Omega_R$) and sufficiently isolated that the bath spectrum does not vary appreciably across the resonance window. The condition that the bath spectrum be slowly-varying near the resonance requires
\begin{equation}
\left|\left.\frac{d\ln J_k(\omega)}{d \omega}\right|_{\omega=\Omega_R}\right|\gamma\ll 1,
\label{J_slowly}
\end{equation} which, using Eq. (\ref{J_fit}), gives $|k|\gamma\ll \Omega_R$. This is obviously satisfied in the weak-damping regime because $|k| \sim \mathcal{O}(1)$. The dissipative response is then dominated by the pole in the experimentally-relevant frequency window, and the corresponding quality factor is
\begin{equation}
\mathcal{Q}\approx \frac{\Omega_R}{\Delta \omega} \approx \frac{\Omega_R}{\gamma}.
\end{equation}
For $\mathcal{Q}\approx 215$ taken in \cite{Groeblacher_2015}, one has $\gamma/\Omega_R \approx 4.65\times 10^{-3}$, justifying the weak-damping approximation as well as the condition in Eq. (\ref{J_slowly}), leading to a slowly-varying bath spectrum over the resonance window. Physically, the quantity $J_k(\Omega_R)$ measures the spectral weight available in the bath at the resonance frequency and therefore controls the efficiency with which the mechanical mode dissipates energy into its environment. A large quality factor corresponds to weak damping, a narrow resonance, and a long-lived dressed mechanical mode, while a larger $\gamma$ implies a broader spectral response and faster relaxation. In this sense, Eq. (\ref{gamma_pole}) provides the link between the near-resonance bath spectrum and the experimentally-accessible sharpness of the resonance within the weak-damping regime. 

\subsection{Dissipation kernel}
\begin{figure}
\centering
\includegraphics[width=\linewidth]{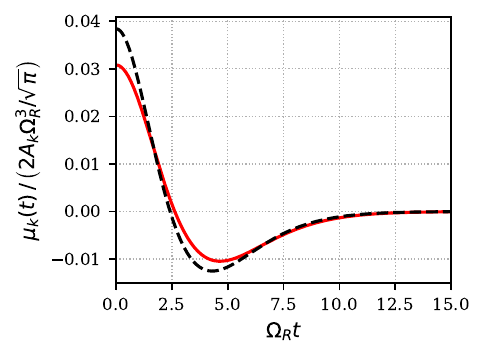}
\caption{\justifying{Normalized dissipation kernel $\mu_k(t)/(2A_k\Omega_R^3/\sqrt{\pi})$ as a function of the dimensionless time $\Omega_R t$. The curves represent $k = -2.30$ (solid-red) and $k = -1.75$ (dashed-black). The asymptotic decay is characterized by a power-law-modulated exponential envelope $\sim t^{2-k} e^{-\Omega_R t}$. The transient sign change reflects delayed, history-dependent friction and provides a time-domain signature of the bath memory.}}
\label{fig3}
\end{figure}

It is of physical interest to investigate the behavior of the dissipation kernel $\mu_k(t)$. Moreover, since $\Omega_R/2\pi = 0.914~{\rm MHz}$ and $T = 300~{\rm K}$ \cite{Groeblacher_2015}, one finds that $\omega \ll k_B T$ (in units where $\hbar = 1$) over the relevant frequency window. As a result, over the frequency range carrying appreciable spectral weight, the noise correlation [Eq. (\ref{noise_correlations})] gives
\begin{equation}
C_{FF,k}(t) \approx k_B T \mu_k(t),
\end{equation} upon using $\coth (\omega/2k_B T) \approx 2k_B T/\omega$. In other words, in the high-temperature limit of this Gaussian model, the noise kernel is proportional to the dissipation kernel, meaning that the relevant temporal-memory structure is encoded within the kernel $\mu_k(t)$. Substituting Eq. (\ref{J_fit}) into Eq. (\ref{mut_J}) gives
\begin{equation}
\mu_k(t)= \frac{2A_k\Omega_R^{2(3-k)}}{\pi} \int_0^\infty \frac{\omega^2\cos(\omega t)}{(\omega^2+\Omega_R^2)^{3-k}}d\omega.
\label{mu_upto_integral}
\end{equation}  
The cosine transform can be evaluated exactly to yield the analytical result (see Appendix \ref{appD})
\begin{equation}
\mu_k(t)= \frac{2A_k\Omega_R^{6-2k}}{\sqrt{\pi}} \left[B_k(t) - D_k(t)\right],
\label{mu_final_form}
\end{equation}
where
\begin{eqnarray}
B_k(t) &=& \frac{\left(\frac{t}{2\Omega_R}\right)^{\frac{3}{2}-k}}{\Gamma(2-k)} K_{\frac{3}{2}-k}(\Omega_R t), \\
D_k(t) &=& \Omega_R^2 \frac{\left(\frac{t}{2\Omega_R}\right)^{\frac{5}{2}-k}}{\Gamma(3-k)} K_{\frac{5}{2}-k}(\Omega_R t),
\end{eqnarray} and $t \geq 0$. In the above-mentioned expressions, $K_\nu$ is a modified Bessel function of the second kind. In Fig. \ref{fig3}, we have plotted the behavior of $\mu_k(t)$ for two values of $k$. It depicts how the resonator that experiences non-Markovian dissipation carries the memory of its past dynamics. In particular, in the long-time regime, the asymptotic scaling $K_\nu(z) \sim \sqrt{\pi/2z} e^{-z} (1 + \mathcal{O}(z^{-1}))$ implies
\begin{equation}\label{muk_algebraic_final}
\mu_k(t) \simeq - d_k t^{2-k} e^{-\Omega_R t},
\end{equation} where the coefficient $d_k$ arises from $D_k$, the latter's long-time behavior dominating over that of $B_k$.
The kernel therefore decays exponentially over the timescale set by the inverse resonance frequency $\Omega_R^{-1}$, multiplied by a power law. As depicted in Fig. \ref{fig3}, the memory kernel changes sign and develops a negative tail before ultimately decaying to zero. This establishes nonlocal-in-time, sign-changing retarded friction within the generalized-Langevin description and may be interpreted as a delayed, history-dependent partial cancellation of damping generated by the structured environment. 

\section{Position and momentum correlations}\label{corr_sec}
It is now natural to examine the stationary two-time correlation functions of the resonator. Since our treatment is limited to thermal equilibrium, the symmetrized position correlation function $C_{QQ}(t)=\frac{1}{2}\langle \{Q(t),Q(0)\}\rangle$ takes the form \cite{Weiss_2021}
\begin{equation}
C_{QQ}(t) = \frac{1}{\pi}\int_{0}^{\infty} d\omega \coth\left(\frac{\omega}{2k_B T}\right) {\rm Im}[\chi(\omega)]\cos(\omega t),
\label{CQQ_main}
\end{equation}
where $\chi(\omega)$ is the mechanical susceptibility and $t\geq0$. The corresponding correlation function for the momentum operator, i.e., $C_{PP}(t)=\frac{1}{2}\langle \{P(t),P(0)\}\rangle$, for $t\geq0$ follows from taking $P(t)=M\dot{Q}(t)$, leading to
\begin{equation}
C_{PP}(t) = \frac{M^2}{\pi}\int_{0}^{\infty} d\omega 
\omega^2 \coth\left(\frac{\omega}{2k_B T}\right) {\rm Im}[\chi(\omega)]\cos(\omega t).
\label{CPP_main}
\end{equation}
The above expressions assume $\langle Q \rangle = 0 = \langle P \rangle$, so that at $t=0$ they reduce to the variances discussed in Appendix \ref{appB}. In the resonance-dominated window, the susceptibility is controlled by an isolated weakly-damped pole near the observed mechanical resonance $\Omega_R$. In this regime, using the local form of the susceptibility [Eq. (\ref{chi_eff})] together with the linewidth [Eq. (\ref{gamma_pole})], one has
\begin{equation}
{\rm Im}[\chi_{\rm eff}(\omega)] \approx \frac{\Omega_R\gamma/M_R}{(\Omega_R^2-\omega^2)^2+\Omega_R^2\gamma^2}.
\end{equation}
Substituting this into Eqs. (\ref{CQQ_main}) and (\ref{CPP_main}), and using the weak-damping condition $\gamma \ll \Omega_R$ gives, at the leading order, the following results:
\begin{eqnarray}
C^{\rm pole}_{QQ}(t) &\approx& \frac{2n_R+1}{2M_R\Omega_R} e^{-\gamma t/2}\cos(\Omega_R t),
\label{C_QQ_NR} \\
C^{\rm pole}_{PP}(t) &\approx& \frac{M^2\Omega_R}{2M_R}(2n_R+1) e^{-\gamma t/2}\cos(\Omega_R t),
\label{C_PP_NR}
\end{eqnarray}
where
\begin{equation}
n_R=\left[\exp\left(\frac{\Omega_R}{k_BT}\right)-1\right]^{-1}.
\end{equation}
It may be re-emphasized that here we have taken $P(t)=M\dot{Q}(t)$ to be the canonical momentum appearing in the microscopic Hamiltonian, not $M_R\dot{Q}(t)$, which is why the pole estimate for $C_{PP}(t)$ contains the factor $M^2/M_R$. These expressions describe the experimentally-accessible narrow-band response of the mechanical mode, i.e., the oscillation frequency is set by the observed resonance frequency $\Omega_R$ and the decay envelope is controlled by the linewidth $\gamma$.

\vspace{2mm}

Generally, in the experimentally-relevant high-temperature regime, where $\omega \ll k_B T$ over the relevant frequency scale, one may use $\coth (\omega/2k_B T) \approx 2k_B T/\omega$, so that
\begin{eqnarray}
C_{QQ}(t) &\approx& \frac{2k_B T}{\pi}\int_0^\infty d\omega \frac{{\rm Im}[\chi(\omega)]}{\omega}\cos(\omega t),
\label{CQQ_hightemp} \\
C_{PP}(t) &\approx& \frac{2M^2k_B T}{\pi}\int_0^\infty d\omega \omega {\rm Im}[\chi(\omega)]\cos(\omega t).
\label{CPP_hightemp}
\end{eqnarray}
In the weak-damping regime, the dominant contribution to the full correlation functions comes from the isolated poles of the susceptibility near the observed resonance $\Omega_R$, giving Eqs. (\ref{C_QQ_NR}) and (\ref{C_PP_NR}). However, because the bath spectrum is structured, the correlation functions are not exhausted by this pole contribution alone. To illustrate this, let us consider the position correlation function 
\begin{equation}
C_{QQ}(t) \approx \frac{2k_B T}{\pi}\int_0^\infty d\omega 
\frac{J_k(\omega)}{\omega|\chi^{-1}(\omega)|^2}\cos(\omega t),
\end{equation} obtained from Eq. (\ref{CQQ_hightemp}) by putting ${\rm Im}[\chi(\omega)]=J_k(\omega)/|\chi^{-1}(\omega)|^2$. Writing $C_{QQ}(t)=C_{QQ}^{\rm pole}(t)+\delta C_{QQ}(t)$ formally defines the exact nonpole contribution $\delta C_{QQ}(t)$, whose evaluation requires an explicit subtraction of a suitably-windowed pole contribution from the full spectral integrand. We shall not attempt such a controlled subtraction here. Instead, only to illustrate a possible kernel-controlled background, let us introduce the separate heuristic quantity
\begin{eqnarray}
C_{QQ}^{{\rm bg},{\rm heur}}(t)
&\equiv&
\frac{2k_B T}{\pi D_*}\int_0^\infty d\omega \frac{J_k(\omega)}{\omega}\cos(\omega t)
\nonumber\\
&=&
\frac{k_B T}{D_*}\mu_k(t),
\label{CQQ_background_heuristic}
\end{eqnarray}
where $D_*\sim M_R^2\Omega_R^4$ represents a slowly-varying off-resonant mechanical scale. Eq. (\ref{CQQ_background_heuristic}) is not a controlled approximation to the complete $\delta C_{QQ}(t)$; it only demonstrates that a background component generated by the structured bath may inherit the temporal profile of the dissipation kernel. Physically, in the resonance-dominated window the oscillator behaves as a weakly-damped dressed mode with a sharply-defined frequency and linewidth, so the leading contribution to the correlation functions takes the standard damped-harmonic form. The structured bath does not replace this leading behavior but generates subleading non-Markovian corrections to it. These corrections arise from the nontrivial analytic structure of the bath response and lead to a structured temporal profile of the correlation functions beyond the pole approximation. The coexistence of a dominant resonance contribution and subleading memory-sensitive corrections is a direct consequence of the fact that the present spectral density reproduces the observed near-resonance behavior while remaining globally well defined yet structured.

\section{Nonuniqueness of the global completion}
\label{nonunique_sec}

The locally-inferred, logarithmic slope of the bath spectrum near the mechanical resonance does not uniquely determine the bath spectral density over all frequencies. The spectral density in Eq. (\ref{J_fit}) should therefore not be interpreted as a unique reconstruction of the reservoir but only an analytically-tractable completion of the measured near-resonance behavior. In this section we will make this point explicit by comparing Eq. (\ref{J_fit}) with a power-law spectrum with an exponential cutoff \cite{Weiss_2021,Leggett_1987}. Let us consider the family
\begin{equation}
J_{\rm exp}(\omega)=A_{\rm exp}\omega^s e^{-\omega/\omega_c},
\label{J_exp_general}
\end{equation}
where $s$ controls the infrared behavior and $\omega_c$ is a high-frequency cutoff. The local logarithmic slope of this spectrum is
\begin{equation}
\frac{d\ln J_{\rm exp}}{d\ln\omega}
=
s-\frac{\omega}{\omega_c}.
\end{equation}
Requiring this slope to agree with the experimentally-inferred value $k$ at the observed mechanical resonance $\Omega_R$ immediately requires one to fix $\omega_c=\frac{\Omega_R}{s-k}$. For a direct comparison with Eq. (\ref{J_fit}), let us choose the same minimal-integer infrared exponent $s=3$, which is sufficient to ensure the finiteness of both the stiffness and mass renormalizations. The matched exponential-cutoff spectrum is therefore
\begin{equation}
J_{{\rm exp},k}(\omega)
=
A_{{\rm exp},k}\omega^3
\exp\left[-(3-k)\frac{\omega}{\Omega_R}\right].
\label{J_exp_k}
\end{equation}
For the central estimate $k=-2.30$, one obtains $\omega_c \approx 0.189\Omega_R$, i.e., the cutoff frequency lies much below the observed resonance. To compare the two completions on equal footing, let us normalize them to the same on-resonance spectral weight $J_{{\rm exp},k}(\Omega_R)=J_k(\Omega_R)\equiv J_{k,R}$, thereby fixing
\begin{equation}
A_{{\rm exp},k} = \frac{J_{k,R}}{\Omega_R^3}e^{3-k}.
\label{A_exp}
\end{equation}
The exponential model is then positive, behaves as $J_{{\rm exp},k}(\omega)\sim \omega^3$ for low frequencies, and has an exponentially-suppressed ultraviolet tail, thereby satisfying the same admissibility requirements as Eq. (\ref{J_fit}). Thus the local slope $k$, the positivity of the spectral function, and the finiteness of the renormalization integrals do not uniquely select Eq. (\ref{J_fit}) but define a class of admissible global completions.

\vspace{2mm}

The two spectra also share the same leading near-resonance dissipative physics. In the weak-damping regime, the linewidth of the dressed mechanical pole is controlled by the on-resonance spectral weight in the manner made explicit in Eq. (\ref{gamma_pole}), up to small differences in the near-resonance effective mass $M_R$, provided that the off-resonant spectral weight is not parametrically enhanced. Consequently, once the two spectra are matched at $\Omega_R$, their dominant near-resonance pole structures are the same to leading order, provided the corresponding effective masses $M_R$ are matched or differ only perturbatively. In particular, the resonance-dominated parts of the stationary correlations take the form suggested in Eqs. (\ref{C_QQ_NR}) and (\ref{C_PP_NR}) for both the spectral functions. Hence the leading near-resonance consequences of the model are robust under changes of the global spectral completion, provided the admissible spectra share the same $J_{k,R}$ and the same local logarithmic slope $k$.

\vspace{2mm}

The distinction between admissible completions appears in quantities that probe the off-resonant analytic structure of the bath. One such quantity is the dissipation kernel. For the exponential model, substituting Eq. (\ref{J_exp_k}) into Eq. (\ref{mut_J}) gives
\begin{equation}
\mu_{{\rm exp},k}(t)
=
\frac{4A_{{\rm exp},k}}{\pi}
\frac{\alpha_k(\alpha_k^2-3t^2)}{(\alpha_k^2+t^2)^3},
\label{mu_exp_final}
\end{equation}
where $\alpha_k=(3-k)/\Omega_R$ has dimensions of time. This kernel is explicitly nonlocal in time and changes sign at $t = \alpha_k/\sqrt{3} = (3-k)/(\sqrt{3} \Omega_R)$. Thus a transient negative component of the memory kernel is not exclusive to the algebraic completion in Eq. (\ref{J_fit}), as it also appears in the matched power-law completion with exponential cutoff of Eq. (\ref{J_exp_k}). The long-time envelope, however, is completion-dependent. From Eq. (\ref{mu_exp_final}) one obtains $\mu_{{\rm exp},k}(t) \sim t^{-4}$ for $\Omega_R t \gg 1$, distinct from the long-time behavior in Eq. (\ref{muk_algebraic_final}).  So while the two admissible completions agree in producing nonlocal dissipation and a delayed negative-memory component, they differ qualitatively in their asymptotic tails; Eq. (\ref{J_fit}) gives a power-law-modulated exponential decay controlled by $\Omega_R^{-1}$, whereas the exponential spectrum gives an algebraic tail.

\vspace{2mm}

This comparison clarifies the status of Eq. (\ref{J_fit}). The experimentally-inferred local exponent $k$, together with the convergence of the renormalization integrals, does not determine a unique global spectral density. These conditions only define a class of admissible structured reservoirs. Within this class, the leading near-resonance pole physics is robust in the weak-damping regime because the linewidth depends only on $J_k(\Omega_R)$ and the effective inertia $M_R$, as shown in Eq. (\ref{gamma_pole}), while the local slope controls corrections associated with the variation of the bath spectrum across the linewidth through Eq. (\ref{J_slowly}). In contrast, the precise global spectral maximum, the detailed sign structure of the dissipation kernel, and the asymptotic decay of the memory are completion-dependent. Nevertheless, the usefulness of Eq. (\ref{J_fit}) lies in providing a resonance-anchored, analytically-controlled representative of this admissible class.

\section{Bath spectroscopy}\label{spec_sec}
We shall now perform a theoretical analysis of the optomechanical readout of the nonlocal mechanical response in the weak-coupling (probe) regime. For simplicity we will assume a one-sided cavity with no intrinsic losses. Let us consider a cavity mode $a$ coupled dispersively to the mechanical coordinate $Q$ via the radiation-pressure interaction $\sim G Q a^\dagger a$. In a frame rotating with the laser frequency, the optomechanical Hamiltonian reads
\begin{equation}
H_{\rm om} = H_{\rm m} - \Delta_0 a^\dagger a + G Q a^\dagger a
+ i \sqrt{\kappa}\left(\epsilon_{\rm in} a^\dagger - \epsilon_{\rm in}^\ast a \right),
\end{equation}
where $H_{\rm m}$ is the mechanical Hamiltonian, $\kappa$ is the cavity damping, $\epsilon_{\rm in}$ signifies the strength of the coherent optical drive, and $\Delta_0$ is the cavity detuning. The optical input noise will be included later in the quantum Langevin equation. Let us now expand around the classical steady state in the manner
\begin{equation}
a(t)=\alpha_s+\delta a(t), \quad \quad Q(t)=Q_s+\delta Q(t),
\end{equation}
where $\alpha_s$ and $Q_s$ are the mean intracavity amplitude and mean displacement, respectively. Retaining only the terms linear in fluctuations yields the following linearized Langevin equation for the cavity fluctuation $\delta a$  \cite{Aspelmeyer_2014}:
\begin{equation}
\dot{\delta a}(t)=\left(i\Delta-\frac{\kappa}{2}\right)\delta a(t)-ig\delta Q(t) +\sqrt{\kappa} a_{\rm in}(t),
\label{optical_langevin}
\end{equation}
where $\Delta = \Delta_0 - G Q_s$, and $g = G \alpha_s$ is the linearized optomechanical coupling, taken real by phase choice. In the above, $a_{\rm in}(t)$ denotes the optical input fluctuations. In Fourier space with our convention $\tilde{o}(\omega)=\int_{-\infty}^{\infty}dt  o(t)e^{i\omega t}$, Eq. (\ref{optical_langevin}) can be solved to give
\begin{equation}
\delta\tilde{a}(\omega) = \chi_c(\omega)\left[-ig \delta \tilde{Q}(\omega) +\sqrt{\kappa} \tilde{a}_{\rm in}(\omega)\right]
\end{equation}
for the cavity susceptibility $\chi_c^{-1}(\omega) = \kappa/2-i(\Delta+\omega)$. The mechanical displacement obeys
\begin{equation}
\delta \tilde{Q}(\omega)=\chi(\omega) \tilde{F}_{\rm tot}(\omega),
\label{Q_general_response}
\end{equation}
where the mechanical susceptibility $\chi(\omega)$ is given in Eq. (\ref{chi_ren}). The total force $\tilde{F}_{\rm tot}(\omega)$ generically contains the bath-induced noise, any externally-applied mechanical force, and the radiation-pressure force generated by the optical probe. Since $Q$ is
a dimensional displacement, the coefficient $g=G\alpha_s$ appearing in Eq. (\ref{optical_langevin}) has force-like units. It is useful to introduce the zero-point displacement
$x_{\rm zp}=(2M\Omega_R)^{-1/2}$ and the corresponding linearized coupling rate $g_{\rm lin}=g x_{\rm zp}$. Eliminating the cavity field produces the optical self-energy
\begin{equation}
\Sigma_{\rm opt}(\omega)
=
ig^2\left[\chi_c(\omega)-\chi_c^*(-\omega)\right],
\end{equation}
so that the optically-dressed inverse mechanical susceptibility is $\chi^{-1}(\omega)-\Sigma_{\rm opt}(\omega)$. The weak-probe approximation therefore requires
\begin{equation}
|\Sigma_{\rm opt}(\omega)|
\ll
|\chi^{-1}(\omega)|
\label{weak_probe_condition}
\end{equation}
throughout the reconstruction window. Near a resolved red-sideband resonance, where the counter-rotating cavity response is negligible, a convenient sufficient form is $(4M/M_R)(g_{\rm lin}^2/\kappa\gamma) \ll 1$. Under this condition, the optical spring and optical damping may be neglected to the leading order. In the absence of an external mechanical drive, one then has $\tilde{F}_{\rm tot}(\omega)\approx\tilde{F}(\omega)$ and
$\delta\tilde{Q}(\omega)\approx\chi(\omega)\tilde{F}(\omega)$.

\vspace{2mm}

The output field satisfies the input-output condition \cite{Gardiner_1985}
\begin{equation}
\tilde{a}_{\rm out}(\omega)=\tilde{a}_{\rm in}(\omega)-\sqrt{\kappa} \delta \tilde{a}(\omega).
\end{equation}
One thus immediately finds $\tilde{a}_{\rm out}(\omega) =
\tilde{a}_{\rm in}(\omega) -\sqrt{\kappa}\chi_c(\omega)
[-ig \chi(\omega)\tilde{F}(\omega)+\sqrt{\kappa} \tilde{a}_{\rm in}(\omega)]$, from which the mechanically-induced component reads
\begin{equation}
\tilde{a}_{\rm out}^{({\rm mech})}(\omega) = i\sqrt{\kappa} g \chi_c(\omega)\chi(\omega)\tilde{F}(\omega).
\label{aout_mech_induced}
\end{equation}
Using a homodyne detection that measures the output quadrature \cite{Wiseman_1993,Clerk_2010}, the mechanically-induced contribution appears in the form
\begin{equation}
\tilde{X}_{\theta,{\rm out}}^{({\rm mech})}(\omega) = \Lambda_\theta(\omega) \chi(\omega)\tilde{F}(\omega),
\label{Xout_general}
\end{equation}
where the function $\Lambda_\theta(\omega)$ is (see Appendix \ref{appE})
\begin{equation}
\Lambda_\theta(\omega) = i \sqrt{\frac{\kappa}{2}} g \left[ e^{-i\theta} \chi_c(\omega) - e^{i\theta} \chi_c^*(-\omega) \right],
\label{Lambda_theta_expression}
\end{equation}
with $\theta$ being the homodyne angle. By the Wiener-Khinchin theorem \cite{Clerk_2010}, the quadrature power spectrum can be expressed to the leading order in the weak-probe limit as
\begin{equation}
S_{X_\theta X_\theta}(\omega)
\approx |\Lambda_\theta(\omega)|^2 |\chi(\omega)|^2 S_{FF}(\omega) + S_{\rm imp}(\omega),
\label{Sout_passive}
\end{equation}
where $S_{\rm imp}(\omega)$ signifies the imprecision background \cite{Clerk_2010}, and we have neglected radiation-pressure backaction and optomechanical noise in the weak-coupling regime \cite{Aspelmeyer_2014}. At thermal equilibrium and for $\omega>0$, one has $S_{FF}(\omega)=J_k(\omega)\coth[\omega/(2k_BT)]$ and
${\rm Im}[\chi(\omega)]=J_k(\omega)|\chi(\omega)|^2$. Hence a calibrated passive measurement determines ${\rm Im}[\chi(\omega)]$, and recovering ${\rm Re}[\chi(\omega)]$ additionally requires sufficiently-broadband data and a Kramers-Kronig reconstruction. The coherent-force protocol described subsequently instead determines the complex susceptibility directly within the measured frequency window.

\subsection{Near-resonance spectroscopy}
In the weak-damping regime and provided the response is dominated by an isolated narrow resonance, the susceptibility in a small window around the observed resonance $\Omega_R$ is well approximated by the local form suggested in Eq. (\ref{chi_eff}) with the linewidth given by Eq. (\ref{gamma_pole}). Substituting Eq. (\ref{chi_eff}) into Eq. (\ref{Xout_general}) yields the result
\begin{equation}
\tilde{X}_{\theta,{\rm out}}^{({\rm mech})}(\omega) \approx \frac{\Lambda_\theta(\omega)\tilde{F}(\omega)}
{M_R(\Omega_R^2-\omega^2)-iJ_k(\Omega_R)}.
\end{equation}
For a homodyne angle $\theta$ chosen to maximize displacement transduction near the resonance, and provided $\Lambda_\theta(\omega)$ varies slowly across the mechanical linewidth, one may approximate
$\Lambda_\theta(\omega)\approx \Lambda_\theta(\Omega_R)$. Then the measured optical response inherits the resonance denominator of the mechanical susceptibility as
\begin{equation}
\tilde{X}_{\theta,{\rm out}}^{({\rm mech})}(\omega) \approx
\frac{\Lambda_\theta(\Omega_R)\tilde{F}(\omega)}{M_R(\Omega_R^2-\omega^2)-iJ_k(\Omega_R)}.
\end{equation}
The measured homodyne quadrature therefore functions as a near-resonance probe of the mechanics. Using Eq. (\ref{Sout_passive}), one can infer $|\chi_{\rm eff}(\omega)|^2$ if $|\Lambda_\theta(\Omega_R)|^2$ and $S_{FF}(\omega)$ have been calibrated. 

\subsection{General-frequency reconstruction}
One can, in principle, go beyond this to reconstruct the full $\chi(\omega)$ denominator. For this, it is important to carefully distinguish between the passive spectroscopy discussed so far and the spectroscopy performed when a known coherent forcing is applied to the resonator. To this end, let us apply a calibrated coherent force $F_{\rm ext}(t)$ to the mechanical resonator. Let $\tilde{X}^{\rm (mech,coh)}_{\theta,{\rm out}}(\omega) = \langle \tilde{X}^{\rm (mech)}_{\theta,{\rm out}}(\omega) \rangle $ be the component of the homodyne signal that is phase-locked to the applied drive. In the weak-coupling regime, one can write $\tilde{X}^{\rm (mech,coh)}_{\theta,{\rm out}}(\omega) = \Lambda_\theta(\omega)\chi(\omega)\tilde{F}_{\rm ext}(\omega)$, since $\langle \tilde{F}(\omega) \rangle = 0$ for the thermal noise, allowing us to express
\begin{equation}
\chi(\omega) = \frac{\tilde{X}^{\rm (mech,coh)}_{\theta,{\rm out}}(\omega)}{\Lambda_\theta(\omega)\tilde{F}_{\rm ext}(\omega)}.
\label{chi_readout_general}
\end{equation} 
A direct reconstruction of the complex susceptibility $\chi(\omega)$ is then plausible in principle, provided $\Lambda_\theta(\omega)$ and $\tilde{F}_{\rm ext}(\omega)$ are calibrated, $\Lambda_\theta(\omega)\neq 0$, and probe-induced backaction remains negligible. Once $\chi(\omega)$ is known, one may equivalently reconstruct its reciprocal $\chi^{-1}(\omega)$, thereby gaining access to both the dispersive and dissipative components of the bath self-energy. The principal advantage of expressing the readout in terms of $\chi(\omega)$ is that no near-resonance approximation is required. Combining Eq. (\ref{chi_ren}) with Eq. (\ref{chi_readout_general}), one finds
\begin{equation}
M\Omega_0^2-M\omega^2-\delta K-\Sigma(\omega) = \frac{\Lambda_\theta(\omega)\tilde{F}_{\rm ext}(\omega)}
{\tilde{X}^{\rm (mech,coh)}_{\theta,{\rm out}}(\omega)}.
\end{equation}
Equivalently, separating the real and imaginary parts, one can write
\begin{equation}
{\rm Re} [\Sigma(\omega)] = M\Omega_0^2-M\omega^2-\delta K
- {\rm Re}\left[\frac{\Lambda_\theta(\omega)\tilde{F}_{\rm ext}(\omega)}{\tilde{X}^{\rm (mech,coh)}_{\theta,{\rm out}}(\omega)}\right],
\end{equation}
and
\begin{equation}
J_k(\omega) = {\rm Im}[\Sigma(\omega)] = -{\rm Im}\left[
\frac{\Lambda_\theta(\omega)\tilde{F}_{\rm ext}(\omega)}
{\tilde{X}^{\rm (mech,coh)}_{\theta,{\rm out}}(\omega)}\right].
\end{equation}
As noted earlier, these reconstruction formulas assume linear response, a calibrated $\Lambda_\theta(\omega)$, and negligible optomechanical backaction on the intrinsic mechanical susceptibility. In the present formulation, the optomechanical formalism is seen to permit, at least in principle, a reconstruction of the frequency-dependent complex susceptibility. The corresponding bath self-energy can then be inferred once the bare mechanical parameters and the chosen renormalization convention are fixed independently. Such a reconstruction directly separates the dispersive renormalization ${\rm Re}[\Sigma(\omega)]$ from the dissipative spectral weight $J_k(\omega)$ within the measured frequency window. On the other hand, at thermal equilibrium, passive spectroscopy determines ${\rm Im}[\chi(\omega)]$ through the fluctuation-dissipation relation following Eq. (\ref{Sout_passive}) and can, in principle, recover ${\rm Re}[\chi(\omega)]$ through a broadband Kramers-Kronig reconstruction. The advantage of coherent-force spectroscopy is that the complex susceptibility is obtained directly from the phase-sensitive response, without requiring this global reconstruction.

\vspace{2mm}

It should be strongly emphasized that the present approach using coherent-force signal is distinct from the optical-transmission-based spectral-density reconstruction reported in \cite{Zhang_2021}. While in the earlier approach the measured optical transmission is used to infer the reservoir spectral density, the present coherent-force spectroscopy is formulated directly at the level of the complex mechanical susceptibility. Because the applied force is calibrated and the homodyne response is phase-sensitive, the protocol retains the phase of the mechanical response and therefore permits, in principle, the reconstruction of the full inverse susceptibility. This gives access not only to the dissipative component $J(\omega)={\rm Im}[\Sigma(\omega)]$, but also to the frequency-dependent dispersive part ${\rm Re}[\Sigma(\omega)]$.

\subsection{Numerical reconstruction from simulated coherent-force data}\label{num_reconstruction_sec}

Let us now present the result of a compact simulation of the coherent-force reconstruction protocol, aimed at showing how the complex bath self-energy can be recovered from a calibrated phase-sensitive optomechanical response. For this purpose we shall use the model spectral density in Eq. (\ref{J_fit}) with $k=-2.30$, set $\Omega_R=1$ and $M=1$, and take $\mathcal{Q}=215$. Figs. \ref{fig4} and \ref{fig5} show the resulting reconstruction (see Appendix \ref{appF} for the details). The black curves are the exact input model, and the red points are obtained from the noisy and slightly-miscalibrated coherent response. The imaginary part reproduces the bath spectral function, while the real part captures the frequency-dependent dispersive correction. This demonstrates the additional information obtained from coherent-force spectroscopy: once the complex susceptibility $\chi(\omega)$ is reconstructed, the dissipative and dispersive bath contributions can be separated. Compared with passive finite-band noise spectroscopy, the coherent-force protocol directly retains the phase of the response and reconstructs $\chi(\omega)$ pointwise. 

\vspace{2mm}

The reconstruction accuracy is limited by three main factors. First, the coherent response must remain above the imprecision floor over the frequency window of interest. Second, over the finite reconstruction window considered here, the dispersive and dissipative components are affected predominantly by amplitude- and phase-calibration errors, respectively, because amplitude errors rescale the inverse susceptibility, whereas phase errors rotate and mix its real and imaginary parts. Third, the weak-probe condition must be maintained, for otherwise the reconstructed susceptibility corresponds to an optically-dressed mechanical response and the optical self-energy must be independently calibrated and subtracted.

\begin{figure}
\centering
\includegraphics[width=\linewidth]{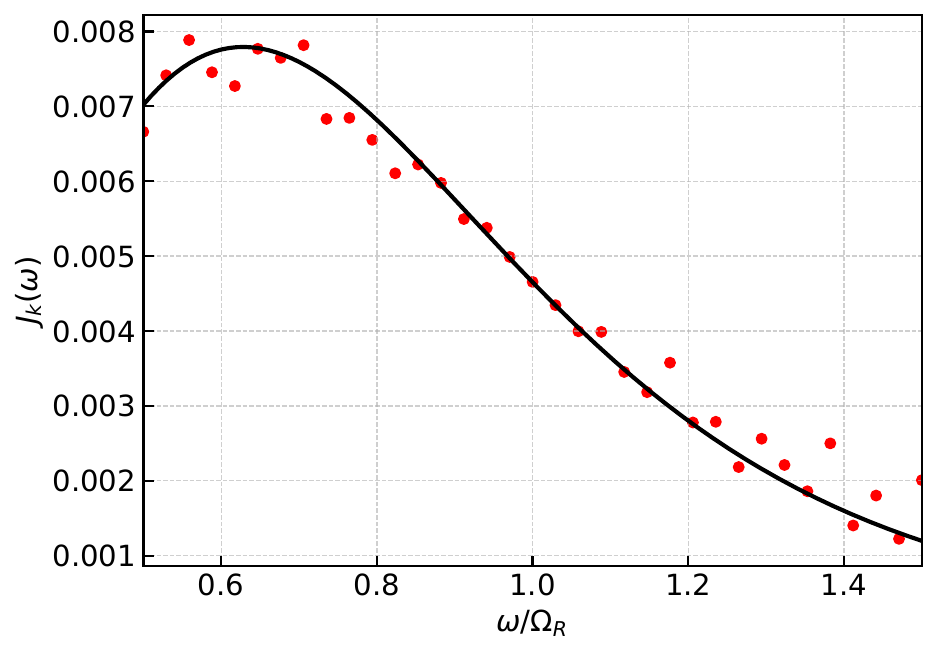}
\caption{\justifying{Numerical reconstruction of the dissipative component of the bath self-energy from calibrated coherent-force spectroscopy. The black curve shows the exact model spectral density $J_k(\omega)$ from Eq. (\ref{J_fit}) for $k=-2.30$, while the red points show the reconstructed values $J_{k,{\rm rec}}(\omega)={\rm Im}[\Sigma_{\rm rec}(\omega)]$ obtained from the noisy coherent homodyne response. The simulation uses $\Omega_R=1$, $M=1$, $\mathcal{Q}=215$, $\kappa=0.2\Omega_R$, $\Delta=-\Omega_R$, $g_{\rm lin}=g x_{\rm zp}=10^{-3}\Omega_R$, and $\theta=\pi/2$, with a signal-to-noise ratio of $60{\rm ~dB}$ and small amplitude and phase calibration errors. The qualitative agreement demonstrates that the spectral density can be recovered from the reconstructed complex susceptibility under calibrated weak-probe conditions.}}
\label{fig4}
\end{figure}

\begin{figure}
\centering
\includegraphics[width=\linewidth]{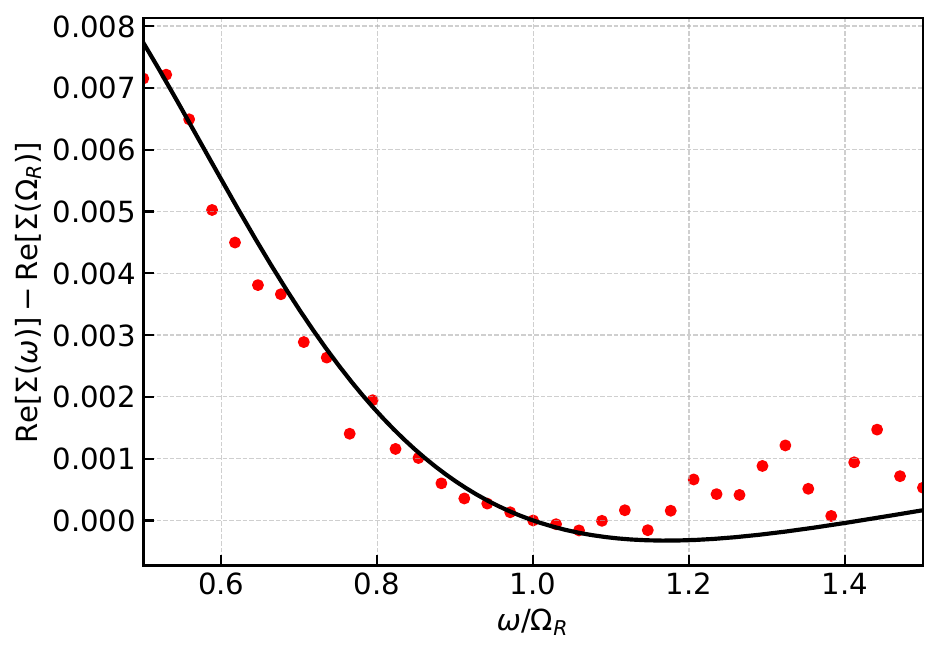}
\caption{\justifying{Numerical reconstruction of the dispersive component of the bath self-energy from the same calibrated coherent-force measurement as in Fig. \ref{fig4}. The black curve shows the exact offset-free dispersive correction $\Delta{\rm Re}[\Sigma(\omega)]={\rm Re}[\Sigma(\omega)]-{\rm Re}[\Sigma(\Omega_R)]$, while the red points show the reconstructed values $\Delta{\rm Re}[\Sigma_{\rm rec}(\omega)]={\rm Re}[\Sigma_{\rm rec}(\omega)]-{\rm Re}[\Sigma_{\rm rec}(\Omega_R)]$. The offset-free quantity is shown because the absolute value of ${\rm Re}[\Sigma(\omega)]$ depends on the bare-frequency convention. The qualitative agreement confirms that coherent-force spectroscopy can access the frequency-dependent dispersive bath response under calibrated weak-probe conditions.}}
\label{fig5}
\end{figure}

\section{Conclusions}\label{conc_sec}
In this work we have theoretically developed a consistent framework for describing the non-Markovian Brownian motion of an optomechanical resonator, explicitly incorporating the experimentally-observed non-Ohmic behavior near the mechanical resonance. Recognizing that a global extrapolation of the locally-inferred spectral scaling leads to unphysical
divergences, we constructed a phenomenological bath spectral density that is globally well defined, while faithfully reproducing the measured near-resonance behavior. This construction ensures finite bath-induced renormalizations of both the effective mass and stiffness, thereby yielding a physically-admissible description of the dressed mechanical dynamics. Within this framework we have shown that the resulting nonlocal mechanical response is governed by a dissipation kernel exhibiting a power-law-modulated exponential decay accompanied by a transient negative regime. These features provide time-domain signatures of sign-changing retarded friction and bath memory within the generalized-Langevin description. As we have argued, the global completion is not unique. In the weak-damping near-resonance regime, the leading response remains accurately described by the local isolated-pole approximation.

\vspace{2mm}

Further, we have analyzed the optomechanical readout in the weak-coupling regime and have demonstrated that homodyne detection, when combined with a calibrated coherent drive, may, in principle, access the full complex mechanical susceptibility. This provides, in principle, a route to separating the dissipative and dispersive components of the bath self-energy under calibrated coherent-force spectroscopy, which we have illustrated using a numerical simulation. Taken together, our results establish a direct link between locally-inferred experimental signatures and globally-consistent, open-system modeling, providing a unified framework for investigating structured environments and memory effects in optomechanics. Beyond its immediate applicability, this approach provides a framework for systematic reservoir engineering and the controlled exploration of non-Markovian dynamics in micromechanical systems.\\

\textbf{Acknowledgements:} We are thankful to the anonymous referees for their constructive comments, which have led to a substantial improvement of the paper. A.G. thanks Gert-Ludwig Ingold for useful discussions on quantum Brownian motion. M.B. thanks the Air Force Office of Scientific Research (AFOSR) (FA9550-23-1-0259) for support.

\appendix

\begin{widetext}

\section{Derivation of quantum Langevin equation and conventions}\label{appA} 
Let us briefly outline the derivation of the quantum Langevin equation [Eq. (\ref{QLE})] starting from the linear-coupling Hamiltonian [Eq. (\ref{H})]. The Heisenberg equations are
\begin{equation}
M\ddot{Q}(t) + M\Omega_0^2 Q(t) = \sum_{j=1}^N c_j x_j(t), \quad \quad \ddot{x}_j(t) + \omega_j^2 x_j(t) = \frac{c_j}{m_j} Q(t).
\end{equation}
Solving the bath equation for the initial time $t_0$ gives
\begin{equation}
x_j(t) = x_j(t_0)\cos[\omega_j(t-t_0)] + \frac{p_j(t_0)}{m_j\omega_j}\sin[\omega_j(t-t_0)] + \int_{t_0}^t \frac{c_j}{m_j\omega_j} \sin[\omega_j(t-t')] Q(t') dt'.
\end{equation} 
The integral term can be evaluated using integration by parts to yield
\begin{equation}
\int_{t_0}^t \sin[\omega_j(t-t')] Q(t') dt' = \frac{Q(t)}{\omega_j} - \frac{Q(t_0)\cos[\omega_j(t-t_0)]}{\omega_j} - \int_{t_0}^t \frac{\cos[\omega_j(t-t')]}{\omega_j} \dot{Q}(t') dt'.
\end{equation}
Now substituting this expanded integral into the system's equation of motion and grouping the terms, one finds
\begin{equation}
M\ddot{Q}(t) + M\Omega_0^2 Q(t) - Q(t) \sum_{j=1}^N \frac{c_j^2}{m_j\omega_j^2} + \int_{t_0}^t \mu(t-t') \dot{Q}(t') dt' = F(t) + ({\rm boundary~term~at~} t_0),
\end{equation}
where
\begin{equation}
\mu(t) = \sum_{j=1}^N \frac{c_j^2}{m_j\omega_j^2} \cos(\omega_j t), \quad \quad F(t) = \sum_{j=1}^N c_j \left[ x_j(t_0)\cos[\omega_j(t-t_0)] + \frac{p_j(t_0)}{m_j\omega_j}\sin[\omega_j(t-t_0)] \right].
\end{equation}
Converting the discrete sum to an integral in the expression for $\mu(t)$ above exactly reproduces Eq. (\ref{mut_J}). Furthermore, evaluating this kernel at $t=0$ immediately recovers the definition of the stiffness shift $\delta K = \mu(0)$ [Eq. (\ref{deltaK_def})], which naturally shifts the bare restoring force to $M\Omega_0^2 - \delta K$. The right-hand side contains the operator-valued thermal noise that depends only on the free evolution of the bath's initial degrees of freedom, and pushing the initial time $t_0 \to -\infty$ eliminates the transient boundary terms.

\vspace{2mm}

To derive the noise correlations, let us assume the uncoupled bath is initially in a thermal-equilibrium state at temperature $T$. One then has the expectation values $\langle x_j \rangle = \langle p_j \rangle = 0$, leading to $\langle F(t) \rangle = 0$. The nonvanishing second moments are
\begin{equation}
\langle x_j x_k \rangle = \delta_{j,k} \frac{1}{2m_j\omega_j} \coth\left(\frac{\omega_j}{2 k_B T}\right), \quad \quad \langle p_j p_k \rangle = \delta_{j,k} \frac{m_j\omega_j}{2} \coth\left(\frac{\omega_j}{2 k_B T}\right),
\end{equation} and using these to calculate the symmetrized correlation function $C_{FF}(t-t') = \frac{1}{2}\langle \{F(t), F(t')\} \rangle$, one gets the result
\begin{equation}
C_{FF}(t-t') = \sum_{j=1}^N \frac{c_j^2}{2m_j\omega_j} \coth\left(\frac{\omega_j}{2 k_B T}\right) \cos[\omega_j(t-t')].
\end{equation}
In the continuum limit of the heat bath, the above-mentioned correlation function coincides with Eq. (\ref{noise_correlations}) of the main text. 

\section{Convergence of the variances of the canonical quadratures}\label{appB}
A physical requirement is that the stationary fluctuations of the resonator must be well defined. At thermal equilibrium with $T > 0$, the position and momentum variances are determined by the fluctuation-dissipation relations \cite{Weiss_2021}
\begin{equation}
\sigma_Q^2 = \frac{1}{\pi}\int_0^\infty d\omega 
\coth\left(\frac{\omega}{2 k_B T}\right) 
{\rm Im}[\chi(\omega)], \quad \quad 
\sigma_P^2 = \frac{M^2}{\pi}\int_0^\infty d\omega 
\omega^2\coth\left(\frac{\omega}{2 k_B T}\right) 
{\rm Im}[\chi(\omega)],
\end{equation}
with $\chi(\omega)$ given in Eq. (\ref{chi_general}). We shall assume that the dressed resonator is stable, i.e., $M\Omega_0^2-\delta K>0$. One has ${\rm Im}[\chi(\omega)] \sim J_k(\omega)$ as $\omega \to 0$, whereas in the high-frequency regime, the inertial term dominates so that ${\rm Im}[\chi(\omega)] \sim J_k(\omega)/\omega^4$ as $\omega\to\infty$. Therefore, if the spectral density scales as $J_k(\omega)=\mathcal{O}(\omega^s)$ for $\omega\to0$ and $J_k(\omega)=\mathcal{O}(\omega^r)$ for $\omega\to\infty$, then finiteness of the variances requires
\begin{eqnarray}
\sigma_Q^2 &<& \infty \quad \Leftarrow \quad s>0,\quad \quad ~~ r<3,\\
\sigma_P^2 &<& \infty \quad \Leftarrow \quad s>-2,\quad \quad r<1.
\end{eqnarray}
A minimal common condition ensuring that both the stationary fluctuations are finite is
\begin{equation}
s>0, \quad \quad r<1.
\end{equation}
These requirements are weaker than the stronger conditions $s>2$ and $r<0$ imposed to ensure finite $\delta K$ and $\delta M$. In particular, the model spectral density in Eq. (\ref{J_fit}) satisfies these fluctuation bounds comfortably, since it behaves as $J_k(\omega)\sim \omega^3$ in the infrared and $J_k(\omega)\sim \omega^{2k-3}$ in the ultraviolet with $k<0$. 

\section{Exact evaluation of $\delta K$ and $\delta M$}\label{appC} 
Substituting Eq. (\ref{J_fit}) into Eqs. (\ref{deltaK_def}) and (\ref{deltaM_def}) gives
\begin{equation}
\delta K =\frac{2A_k}{\pi} \int_0^\infty d\omega \omega^2
\left[1+\left(\frac{\omega}{\Omega_R}\right)^2\right]^{-(3-k)}, \quad \quad 
\delta M = \frac{2A_k}{\pi} \int_0^\infty d\omega \left[1+\left(\frac{\omega}{\Omega_R}\right)^2\right]^{-(3-k)}.
\end{equation}
Setting $u=\omega/\Omega_R$ then yields
\begin{equation}
\delta K = \frac{2A_k\Omega_R^3}{\pi} \int_0^\infty du
u^2(1+u^2)^{-(3-k)}, \quad \quad 
\delta M = \frac{2A_k\Omega_R}{\pi} \int_0^\infty du (1+u^2)^{-(3-k)}.
\end{equation}
Using the beta-function identity $\int_0^\infty v^{\mu-1}(1+v)^{-\nu}dv = \frac{\Gamma(\mu)\Gamma(\nu-\mu)}{\Gamma(\nu)}$, we have the standard integrals (for $v = u^2$)
\begin{equation}
\int_0^\infty du u^2(1+u^2)^{-a} = \frac{\sqrt{\pi}}{4} 
\frac{\Gamma\left(a-\frac32\right)}{\Gamma(a)},
\quad \quad
\int_0^\infty du(1+u^2)^{-b} = \frac{\sqrt{\pi}}{2} 
\frac{\Gamma\left(b-\frac12\right)}{\Gamma(b)},
\end{equation}
with $a>\frac{3}{2}$ and $b>\frac{1}{2}$. By choosing $a = b = 3-k$, one obtains the exact expressions
\begin{equation}
\delta K = \frac{A_k\Omega_R^3}{2\sqrt{\pi}}
\frac{\Gamma\left(\frac32-k\right)}{\Gamma(3-k)}, \quad \quad \delta M = \frac{A_k\Omega_R}{\sqrt{\pi}}
\frac{\Gamma\left(\frac52-k\right)}{\Gamma(3-k)},
\label{deltaM_deltaK_exact}
\end{equation}
which, upon substituting Eq. (\ref{A_k}), gives
\begin{equation}
\delta K = \frac{2^{2-k} J_k(\Omega_R)}{\sqrt{\pi}} \frac{\Gamma\left(\frac32-k\right)}{\Gamma(3-k)}, \quad \quad
\delta M = \frac{2^{3-k}J_k(\Omega_R)}{\Omega_R^2 \sqrt{\pi}} \frac{\Gamma\left(\frac52-k\right)}{\Gamma(3-k)}.
\end{equation} 
The bath-induced stiffness and mass renormalizations of the resonator are therefore determined by the on-resonance value of the spectral function $J_k(\Omega_R)$, within the chosen global extrapolation in Eq. (\ref{J_fit}). Further, Eq. (\ref{pole_condition}) implies
\begin{equation}
\Omega_0^2 = \left(1+\frac{\delta M}{M}\right)\Omega_R^2 + \frac{\delta K}{M} + \frac{{\rm Re}[\Sigma_{\rm res}(\Omega_R)]}{M}.
\label{Omega0_estimation}
\end{equation}
Eq. (\ref{Omega0_estimation}) shows how the bare frequency $\Omega_0$ is related to the observed resonance $\Omega_R$ once the bath-induced renormalizations of the stiffness and mass are taken into account. 

\section{Exact evaluation of the kernel $\mu_k(t)$}\label{appD}
To evaluate the integral of Eq. (\ref{mu_upto_integral}) analytically, let us resort to the algebraic decomposition $\omega^2 = (\omega^2 + \Omega_R^2) - \Omega_R^2$. This allows us to split the integral into two parts that share a common structural form. Explicitly, one has
\begin{equation}
\mu_k(t) = \frac{2A_k\Omega_R^{6-2k}}{\pi} \left[ \int_0^\infty \frac{\cos(\omega t)}{(\omega^2+\Omega_R^2)^{2-k}} d\omega - \Omega_R^2 \int_0^\infty \frac{\cos(\omega t)}{(\omega^2+\Omega_R^2)^{3-k}} d\omega \right].
\label{mu_integral_transform}
\end{equation}
Both the integrals can now be evaluated using the standard identity for the cosine transform of an inverse polynomial power, yielding the modified Bessel function of the second kind as \cite{Gradshteyn_2007}
\begin{equation}
\int_0^\infty \frac{\cos(u t)}{(u^2+\zeta^2)^m} du = \frac{\sqrt{\pi}}{\Gamma(m)} \left(\frac{t}{2\zeta}\right)^{m-1/2} K_{m-1/2}(\zeta t),
\end{equation}
valid for ${\rm Re}[m] > 1/2$, $\zeta > 0$, and real $t$. Applying this identity to the first integral with $m = 2-k$, $u = \omega$, and $\zeta = \Omega_R$ gives us
\begin{equation}
\int_0^\infty \frac{\cos(\omega t)}{(\omega^2+\Omega_R^2)^{2-k}} d\omega = \sqrt{\pi} \frac{\left(\frac{t}{2\Omega_R}\right)^{\frac{3}{2}-k}}{\Gamma(2-k)} K_{\frac{3}{2}-k}(\Omega_R t) = \sqrt{\pi} B_k(t).
\label{K_int_1}
\end{equation}
Similarly, applying the identity to the second integral with $m = 3-k$ gives
\begin{equation}
\int_0^\infty \frac{\cos(\omega t)}{(\omega^2+\Omega_R^2)^{3-k}} d\omega = \frac{\sqrt{\pi}}{\Omega_R^2} \left[ \Omega_R^2 \frac{\left(\frac{t}{2\Omega_R}\right)^{\frac{5}{2}-k}}{\Gamma(3-k)} K_{\frac{5}{2}-k}(\Omega_R t) \right] = \frac{\sqrt{\pi}}{\Omega_R^2} D_k(t).
\label{K_int_2}
\end{equation}
Substituting Eqs. (\ref{K_int_1}) and (\ref{K_int_2}) into Eq. (\ref{mu_integral_transform}) directly leads to the form given in Eq. (\ref{mu_final_form}) of the main text.

\section{Derivation of the expression for $\Lambda_\theta(\omega)$}\label{appE}
The output homodyne quadrature can be defined as \cite{Clerk_2010}
\begin{equation}
X_{\theta,{\rm out}}(t)=\frac{e^{-i\theta}a_{\rm out}(t)+e^{i\theta}a_{\rm out}^\dagger(t)}{\sqrt2}.
\end{equation}
Starting from the mechanically-induced part of the output field given in Eq. (\ref{aout_mech_induced}), using $\widetilde{a^\dagger}(\omega)=\tilde{a}^\dagger(-\omega)$, together with $\chi(\omega) = \chi^*(-\omega)$ and $\tilde{F}(\omega) = \tilde{F}^\dagger(-\omega)$, we can write
\begin{equation}
\tilde{X}_{\theta,{\rm out}}^{({\rm mech})}(\omega) = i\sqrt{\frac{\kappa}{2}}g \left[e^{-i\theta}\chi_c(\omega)-e^{i\theta}\chi_c^*(-\omega)\right]\chi(\omega)\tilde{F}(\omega).
\end{equation}
Now Eq. (\ref{Lambda_theta_expression}) of the main text follows directly.

\section{Numerical-reconstruction details}\label{appF}

Here we shall summarize the details of the numerical reconstruction given in Sec. \ref{num_reconstruction_sec}. For this we have set $M=1$, $\Omega_R=1$, and $\mathcal Q=215$. The target linewidth is therefore $\gamma=\Omega_R/\mathcal Q$. The on-resonance spectral weight is fixed from the weak-damping linewidth estimate by choosing the effective-mass scale in Eq. (\ref{gamma_pole}) to be unity, namely, $J_k(\Omega_R)=\Omega_R \gamma$. This should be understood as a normalization convention used to set the overall strength of the spectral density in the simulated data. The susceptibility itself is generated from the full expression given in Eq. (\ref{chi_general}) with $M=1$, and not from the local approximate form alone. Consequently, the actual local effective mass associated with the generated full susceptibility is given by 
\begin{equation}
M_R^{\rm loc}
=
M+
\left.
\frac{\partial {\rm Re}\Sigma(\omega)}
{\partial(\omega^2)}
\right|_{\omega=\Omega_R},
\end{equation}
which need not be exactly equal to unity. For the parameters used here this correction is small, and so the chosen normalization gives the intended
weak-damping linewidth to good accuracy. The prefactor $A_k$ is then obtained from Eq. (\ref{A_k}), the stiffness shift $\delta K$ is evaluated from Eq. (\ref{deltaM_deltaK_exact}), while the dispersive self-energy is computed numerically from the regularized principal-value expression in Eq. (\ref{principal_value_exp}). The bare frequency $\Omega_0$ is fixed directly from the resonance convention ${\rm Re}[\chi^{-1}(\Omega_R)]=0$. When removing the principal-value singularity by subtracting its value at $\omega'=\omega$, the finite numerical upper cutoff $\omega_{\rm cut}^{(\rm num)}$ is treated explicitly. In particular, one has
\begin{equation}
\mathcal{P}\int_0^{\omega_{\rm cut}^{(\rm num)}}
\frac{d\omega'}{\omega'^2-\omega^2} = \frac{1}{2\omega} \ln\left|\frac{\omega_{\rm cut}^{(\rm num)}-\omega}{\omega_{\rm cut}^{(\rm num)}+\omega}\right|,
\end{equation}
and this finite-cutoff contribution is retained in the numerical evaluation. We have used $\omega_{\rm cut}^{(\rm num)} = 120 \Omega_R$.

\vspace{2mm}

The synthetic homodyne signal is generated from Eq. (\ref{chi_readout_general}), retaining the full optical transduction factor $\Lambda_\theta(\omega)$ in Eq. (\ref{Lambda_theta_expression}). We have taken the sample parameters
\begin{equation}
\kappa=0.2\Omega_R,\quad \quad
\Delta=-\Omega_R,\quad \quad
g_{\rm lin}=g x_{\rm zp}=10^{-3}\Omega_R,\quad \quad \theta=\pi/2.
\end{equation}
For these parameters, $M_R^{\rm loc}$ differs from $M$ only perturbatively and
\begin{equation}
\frac{4M}{M_R^{\rm loc}}
\frac{g_{\rm lin}^2}{\kappa\gamma}
\simeq4.3\times10^{-3}\ll1,
\end{equation}
so the optical self-energy is perturbative and the weak-probe condition in Eq. (\ref{weak_probe_condition}) is satisfied. The reconstruction is performed over the window $0.5\leq\omega/\Omega_R\leq1.5$. To avoid an ill-conditioned inversion in regions where a fixed-amplitude force would give a very small coherent response, we have modeled a calibrated swept-force measurement in which the known complex amplitude of the applied coherent force is chosen frequency by frequency so that the coherent homodyne response has an approximately-constant magnitude across the frequency sweep. The applied force, including its phase, is assumed to be calibrated at every frequency and is explicitly divided out in the reconstruction. In the numerical simulation, this swept-force protocol is implemented using the known model response only to impose a nearly-uniform signal-to-noise ratio across the plotted window. So it does not model actuator-amplitude constraints or establish the absolute force required in a particular experiment, and the overall coherent-force and signal amplitudes may be rescaled together without altering the reconstruction at fixed signal-to-noise ratio. It should be emphasized that the chosen parameters are representative and are intended only to demonstrate proof-of-principle reconstruction.

\vspace{2mm}

Finite measurement imperfections are included by adding complex Gaussian imprecision noise corresponding to a $60~{\rm dB}$ amplitude signal-to-noise ratio, i.e., an amplitude ratio of $10^3$, equivalently a power ratio of $10^6$. Calibration errors are included by replacing the exact transduction factor and force by
\begin{equation}
\Lambda_\theta^{\rm cal}(\omega)
=
\Lambda_\theta(\omega)(1+\epsilon_\Lambda)e^{i\phi_\Lambda},
\quad \quad
F_{\rm ext}^{\rm cal}(\omega)
=
F_{\rm ext}(\omega)(1+\epsilon_F)e^{i\phi_F},
\end{equation}
with $\epsilon_\Lambda=5\times10^{-4}$, $\epsilon_F=3\times10^{-4}$, $\phi_\Lambda=0.010^\circ$, and $\phi_F=0.006^\circ$. The finite optical bandwidth is included through the cavity linewidth $\kappa$. The susceptibility is reconstructed using Eq. (\ref{chi_readout_general}), and the self-energy is then obtained by
inverting Eq. (\ref{chi_general}). In our convention the dissipative part is
$J_{k,{\rm rec}}(\omega)={\rm Im}[\Sigma_{\rm rec}(\omega)]$, while for the dispersive component, we have focused on the quantity
\begin{equation}
\Delta{\rm Re}[\Sigma_{\rm rec}(\omega)] =
{\rm Re}[\Sigma_{\rm rec}(\omega)] -
{\rm Re}[\Sigma_{\rm rec}(\Omega_R)],
\end{equation}
since the absolute offset of ${\rm Re}[\Sigma(\omega)]$ depends on the $\Omega_0$ convention.

\end{widetext}


\begin{thebibliography}{99}

\bibitem{Weiss_2021}
U. Weiss, {\it Quantum dissipative systems}, 5th ed., World Scientific (2021).

\bibitem{Breuer_Petruccione_2002}
H.-P. Breuer and F. Petruccione, {\it The theory of open quantum systems}, Oxford University Press (2002).

\bibitem{Seoanez_2007}
C. Seoanez, F. Guinea, and A. H. Castro Neto, {\it Dissipation due to two-level systems in nano-mechanical devices}, EPL \textbf{78}, 60002 (2007).

\bibitem{Wilson-Rae_2008}
I. Wilson-Rae, {\it Intrinsic dissipation in nanomechanical resonators due to phonon tunneling}, Phys. Rev. B \textbf{77}, 245418 (2008).

\bibitem{Grabert_1988}
H. Grabert, P. Schramm, and G.-L. Ingold, {\it Quantum Brownian motion: The functional integral approach}, Phys. Rep. \textbf{168}, 115 (1988). 

\bibitem{Hanggi_2005}
P. H\"anggi and G.-L. Ingold, {\it Fundamental aspects of quantum Brownian motion}, Chaos \textbf{15}, 026105 (2005).

\bibitem{Ghosh_2024}
A. Ghosh, M. Bandyopadhyay, S. Dattagupta, and S. Gupta, {\it Independent-oscillator model and the quantum Langevin equation for an oscillator: a review}, J. Stat. Mech.: Theory Exp. \textbf{2024}, 074002 (2024).

\bibitem{Kippenberg_2007}
T. J. Kippenberg and K. J. Vahala, {\it Cavity opto-mechanics}, Opt. Exp. \textbf{15}, 17172 (2007).

\bibitem{Aspelmeyer_2014}
M. Aspelmeyer, T. J. Kippenberg, and F. Marquardt, {\it Cavity optomechanics}, Rev. Mod. Phys. \textbf{86}, 1391 (2014).

\bibitem{Groeblacher_2015}
S. Gr{\"o}blacher, A. Trubarov, N. Prigge, G. D. Cole, M. Aspelmeyer, and J. Eisert, {\it Observation of non-Markovian micromechanical Brownian motion}, Nat. Commun. \textbf{6}, 7606 (2015).

\bibitem{Jiang_2020}
W. Jiang, R. Yang, T.-H. Qiu, and G.-J. Yang, {\it Probe response of a cavity-optomechanical system coupling to a frequency-dependent bath}, Phys. Rev. A \textbf{101}, 033804 (2020). 

\bibitem{Zhang_2021}
W.-Z. Zhang, X.-T. Liang, J. Cheng, and L. Zhou, {\it Measurement of the mechanical reservoir spectral density in an optomechanical system}, Phys. Rev. A \textbf{103}, 053707 (2021).

\bibitem{Ullersma_1966}
P. Ullersma, {\it An exactly solvable model for Brownian motion: I. Derivation of the Langevin equation}, Physica \textbf{32}, 27 (1966).

\bibitem{Ford_1988}
G. W. Ford, J. T. Lewis, and R. F. O'Connell, {\it Quantum Langevin equation}, Phys. Rev. A \textbf{37}, 4419 (1988).

\bibitem{Feynman_1963}
R. P. Feynman and F. L. Vernon Jr., {\it The theory of a general quantum system interacting with a linear dissipative system}, Ann. Phys. (N.Y.) \textbf{24}, 118 (1963). 

\bibitem{Ford_1965}
G. W. Ford, M. Kac, and P. Mazur, {\it Statistical mechanics of assemblies of coupled oscillators}, J. Math. Phys. \textbf{6}, 504 (1965).

\bibitem{Caldeira_1983}
A. O. Caldeira and A. J. Leggett, {\it Path integral approach to quantum Brownian motion}, Physica A \textbf{121}, 587 (1983). 

\bibitem{Grabert_1984}
H. Grabert, U. Weiss, and P. Talkner, {\it Quantum theory of the damped harmonic oscillator}, Z. Phys. B \textbf{55}, 87 (1984).

\bibitem{Karrlein_1997}
R. Karrlein and H. Grabert, {\it Exact time evolution and master equations for the damped harmonic oscillator}, Phys. Rev. E \textbf{55}, 153 (1997).

\bibitem{Leggett_1987}
A. J. Leggett, S. Chakravarty, A. T. Dorsey, M. P. A. Fisher, A. Garg, and W. Zwerger, {\it Dynamics of the dissipative two-state system}, Rev. Mod. Phys. \textbf{59}, 1 (1987).

\bibitem{Wiseman_1993}
H. M. Wiseman and G. J. Milburn, {\it Quantum theory of optical feedback via homodyne detection}, Phys. Rev. Lett. \textbf{70}, 548 (1993).

\bibitem{Clerk_2010}
A. A. Clerk, M. H. Devoret, S. M. Girvin, F. Marquardt, and R. J. Schoelkopf, {\it Introduction to quantum noise, measurement, and amplification}, Rev. Mod. Phys. \textbf{82}, 1155 (2010).

\bibitem{Gardiner_1985}
C. W. Gardiner and M. J. Collett, {\it Input and output in damped quantum systems: Quantum stochastic differential equations and the master equation}, Phys. Rev. A \textbf{31}, 3761 (1985).

\bibitem{Gradshteyn_2007}
I. S. Gradshteyn and I. M. Ryzhik, {\it Table of integrals, series, and products}, edited by A. Jeffrey and D. Zwillinger, 7th ed., Academic Press (2007).

\end{thebibliography}
\end{document}